\documentclass[jkps,twocolumn,fleqn,floatfix,showpacs,showkeys]{revtex4}
\usepackage{epsf,amsmath,amssymb,wasysym,bm}
\newcommand{\be}{\begin{equation}}
\newcommand{\ee}{\end{equation}}

\begin{document}
\title{Noise Characteristics of Molecular Oscillations in Simple Genetic Oscillatory Systems}
\author{Byungjoon Min}
\author{K.-I. Goh}
\email{kgoh@korea.ac.kr} 
\author{I.-M. Kim} 
\affiliation{Department of Physics, Korea University, Seoul 136-713, Korea}
\date[]{Received 29 October 2009}
\begin{abstract}
We study the noise characteristics of stochastic oscillations 
in protein number dynamics of simple genetic oscillatory systems.
Using the three-component negative feedback transcription regulatory system
called the repressilator as a prototypical example, 
we quantify the degree of fluctuations in oscillation periods
and amplitudes, as well as the noise propagation along the regulatory
cascade in the stable oscillation regime
via dynamic Monte Carlo simulations.
For the single protein-species level,
the fluctuation in the oscillation amplitudes is found to be
larger than that of the oscillation periods, the distributions of which
are reasonably described by the Weibull distribution and the Gaussian tail, 
respectively. 
Correlations between successive periods and between successive amplitudes,
respectively, are measured to assess the noise propagation properties,
which are found to decay faster for the amplitude than for the period.
The local fluctuation property is also studied.
\end{abstract}
\maketitle
\section{INTRODUCTION}
Molecular events driving cellular processes, such as 
the binding/unbinding of transcription factors to DNA sites,
the transcription of mRNAs, the translation of proteins from
the transcribed mRNAs, and their degradations,
are fundamentally stochastic processes governed by probabilistic
contacts between these molecules and intracellular diffusion \cite{review1}.
Therefore understanding the dynamic fluctuations 
in the genetic regulatory system arising from such stochasticity 
is crucial as they may provide a potential obstacle
in maintaining the robustness of the cellular function 
and the homeostasis of the cellular behavior 
directly or indirectly.
Study on stochastic fluctuations of cellular dynamics
has been accelerated by the recent advances
in synthetic biology \cite{synthetic-review1,synthetic-review2}
and in experimental techniques using the fluorescent proteins
\cite{elowitz-review},
which has allowed us to monitor the fluctuating molecular activities
at the single-cell level in a controlled and designed manner.
Theoretical framework with which stochastic dynamics of
molecule number in time around the steady state can be
analyzed has been developed and applied for simple gene expression 
systems \cite{paulsson1}.
How the fluctuation (or ``noise'') in
one molecular component propagates to the downstream
components in a simple, linear genetic network has also been studied \cite{propa}.

Most studies so far, however, concern 
fluctuations around the steady state, such as
the variance in protein number around its temporally-averaged 
constant mean value.
Meanwhile, there are only a limited number of experimental and theoretical 
studies concerning out-of-steady-state activities, such as 
oscillatory gene expression~\cite{cyano}.
Therefore, their characteristics of dynamic fluctuations are largely unknown.
To address this, here we study the characteristics
of fluctuations in the non-steady states, that of oscillatory 
systems in particular, by using a model genetic oscillatory system.
Undoubtedly, oscillatory system is one of 
the most important systems that never reach steady states,
which underlie a number of important
cellular and physiological processes ranging from bacterial chemotaxis
to mammalian circadian rhythms \cite{goldbeter,cellcycle,stochastic}. 
A number of synthetic gene circuits
with a few components and relatively simple regulatory mechanisms
have been constructed to realize oscillatory dynamics
in protein numbers {\it in vivo} \cite{repressilator,hasty1,mammalian}.
Among these models, we consider the so-called {\em repressilator} (Fig.~1)
\cite{repressilator},
a simple negative feedback circuit with three genetic components,
for its simplicity in the regulatory processes involved, thus making it
appropriate for a starting point in a systematic understanding
of the noise properties of more complicated oscillatory systems.

The repressilator system has been studied theoretically in both
determinisitc \cite{khammash,goh} and stochastic \cite{sasai1,biham1}
frameworks. Most of these works were concerned with the condition and
the stability of sustained oscillations in the circuit with respect to 
various system parameters, such as DNA binding mechanisms, reaction rates,
and plasmid numbers. A general conclusion from these studies
was that the stability and the coherence of the oscillation depends
strongly on the system parameters and that optimal conditions exist
under which the system exhibits the most stable oscillations.
However, even under the optimal conditions, the oscillation is 
never perfect, and ``noise'' is inevitable. It is, thus, necessary
to know the general characteristics of the noise in oscillatory
dynamics in such a stable oscillation regime, which is currently lacking.
Our primary aim in this paper is to fill this gap by performing
a detailed analysis of noisy oscillations in this system.

Fluctuations in stable oscillatory dynamics can be 
classified into two major classes: global and local noise.
The former refers to fluctuations in the
global variables of oscillatory signals, such as 
the oscillation periods and amplitudes, resulting from
a series of temporally organized molecular events.
The latter is mainly due to momentary stochastic stepwise changes
in the molecule number, which occur locally around the overall
oscillatory signal. Previous studies on dynamic fluctuations
have mostly been concerned with the latter.
For the oscillatory activities, however, the characteristics
of global noise or fluctuation is equally, if not more, important,
as both the robust control of periods and achievement of
sufficiently strong and regular strength (amplitudes) of the signal
would be required for an appropriate response of the cell under
fluctuating and noisy condition.
Thus, our focus in this work is toward the characterization
of the global noise and its propagation, with a brief discussion
on the local noise.

This paper is organized as follows: In Sec.~II, we describe the
repressilator model and the methods of numerical calculations.
In Sec.~III, we present the main results and discuss their interpretations
and potential biological meanings. In the final section, we will 
summarize and conclude the paper.

\begin{figure}[t]
\centerline{\epsfxsize=\linewidth \epsfbox{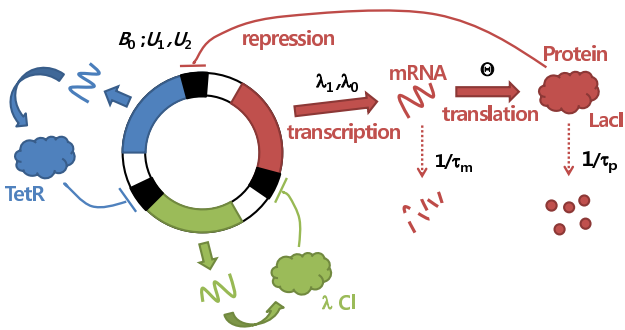}}
\vskip -0.5cm
\caption{ (Color online)
Schematic illustration of the stochastic model of the repressilator
assembled into a plasmid containing three genes, {lacI}, {tetR},
and {$\lambda$-cI}, and its elementary reaction steps. See the text for the details.
}
\end{figure}

\begin{figure*}
\centerline{\epsfxsize=0.9\linewidth \epsfbox{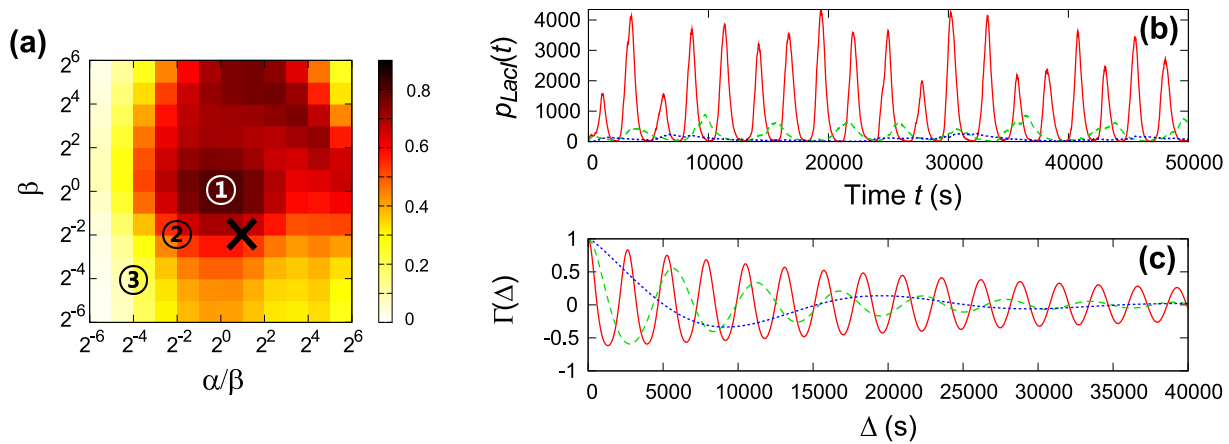}}
\caption{
(Color online) 
{\bf (a)} Stability diagram of the stochastic repressilator system.
Plotted are the color-coded first-peak values of the auto-correlation
function $\Gamma(\Delta)$ of the protein number time series obtained
at each grid point.
{\bf (b)} Typical time-courses of the protein LacI number with time
for three representative parameter sets indicated in (a):
\textcircled{1} (solid), \textcircled{2} (dashed), and \textcircled{3} (dotted).
{\bf (c)} The auto-correlation function of the protein number time series 
shown in (b),
corresponding to the stability measures 0.83(1) (solid), 0.56(2) (dashed), 
and 0.14(3) (dotted). Numbers in parentheses denote the uncertainty in
the last digit.
}
\end{figure*}

\section{Model and Methods}

\subsection{Repressilator Model}
The repressilator \cite{repressilator} is a plasmid fed into an {\it E. coli},
consisting of three transcriptional repressor genes (Fig.~1).
The first gene is lacI from {\it E. coli} itself, 
encoding the transcriptional repressor protein LacI.
LacI inhibits the transcription of the second gene tetR
from the tetracycline-resistance 
transposon Tn10. TetR, the protein product of tetR, 
subsequently inhibits the transcription of the third gene 
cI originating from $\lambda$-phage. 
Finally, its protein product CI inhibits lacI expression,
completing the tri-component negative feedback loop.

Elementary reaction steps involved in this process
include the binding/unbinding of repressor proteins, mRNA transcription,
translation into their protein product, and degradations of
mRNA and protein molecules.
The reaction rate for each of these steps is chosen as a typical
value in the physiological condition following Ref.~\cite{repressilator}: 
The binding rate of repressor proteins to each operator site
is $B_0=1~\textrm{nM}^{-1}\textrm{s}^{-1}$;
the unbinding rate of repressor proteins from the singly-occupied
and doubly-occupied operator sites are 
$U_1=224~\textrm{s}^{-1}$ and $U_2=9~\textrm{s}^{-1}$, respectively;
the protein translation rate is $\Theta=0.167~\textrm{s}^{-1}$ per mRNA molecule;
and the half-life (inverse decay rate) of an mRNA molecule is $\tau_m=120~\textrm{s}$.
The remaining reaction steps are mRNA transcription and protein degradation.
We choose the transcription rate in the free (without inhibitor binding)
state to be $\lambda_1=\alpha~\textrm{s}^{-1}$, and that in the inhibited state
to be much slower as $\lambda_0=10^{-3}\times\lambda_1$.
Finally, the half-life of a protein molecule is set to
be $\tau_p=\tau_m/\beta$ (Fig.~1). 
In this model setting, the two parameters $\alpha$ and $\beta$
control the overall stability of the oscillatory behaviors,
the characteristics of which have been studied in terms of 
deterministic modeling \cite{repressilator,khammash,goh}.

\subsection{Stochastic Simulations: Gillespie Algorithm}
To fully realize the stochastic trajectories of protein number
dynamics, we perform dynamic Monte-Carlo simulations of the reaction processes.
We use the exact stochastic method first developed
by Gillespie in 1970s~\cite{gillespie}.
This algorithm proceeds as follows:
Given a list of reactions that can occur at the moment
$\{R_i\}$ with their reaction rates $\{r_i\}$,
the reaction occurring next is chosen randomly
in proportion to the reaction rates. Then,
the chosen reaction, say $R_k$, occurs, resulting
in changes in the molecule numbers involved in that reaction.
At the same time, the time elapses by $\delta$, which is determined
by a random number sampled from the exponential waiting time
distribution $P_w(\delta)=r_k\exp(-r_k\delta)$, following
from the assumption that the reaction process is
a Poisson process with a rate $r_k$.
We repeat these procedures for a sufficiently long period
of time to obtain a sample trajectory of the given reaction system.

\subsection{Stability of Stochastic Oscillations}
We perform the stochastic simulation for repressilator dynamics
by varying the parameters $\tilde{\alpha}\equiv\alpha/\beta$ and $\beta$,
each from $2^{-10}$ to $2^{10}$.
Depending on the parameter values, the degree of stable oscillations
exhibited by the stochastic dynamics varies \cite{sasai1,biham1}.
To address that, we consider the auto-correlation function 
\begin{equation}
\Gamma_i(\Delta)=\frac{\langle p_i(t)p_i(t+\Delta)\rangle-\langle p_i(t)\rangle
\langle p_i(t+\Delta)\rangle}{\sigma_{p_i(t)}\sigma_{p_i(t+\Delta)}}, 
\end{equation}
as a function of the time lag $\Delta$, 
where $p_i(t)$ is the molecular number of protein $i$ as a function of
time, $\langle x(t)\rangle$ denotes the time average of $x(t)$ over $t$,
and $\sigma_{x(t)}$ is the standard deviation thereof.
For oscillatory signals, we have an oscillatory behavior in $\Gamma(\Delta)$.
For imperfect oscillatory signals, $\Gamma(\Delta)$ decays in amplitude
with $\Delta$ while oscillating.
The more regular the sustained oscillation, 
the slower is the decay in the amplitude of $\Gamma(\Delta)$ with time; 
thus, the first peak in $\Gamma(\Delta)$ $(\Delta>0)$ is higher.
Therefore, we use the value of $\Gamma(\Delta)$ at the first peak 
as a stability measure for stochastic oscillations (Fig.~2).

\subsection{Finding Oscillation Periods and Amplitudes}
Due to constantly fluctuating protein numbers in time,
it is not trivial to pinpoint exact timings of the start and 
the end of each oscillatory cycle under stochastic oscillations.
In this work, we devise an {\it ad-hoc} procedure
to determine the oscillation periods and amplitudes 
from a noisy oscillatory signal by assigning 
top and bottom points of the oscillation in the time series.
To this end, we look at protein number dynamics at a coarse-grained level
by introducing a time window, inside which local variations 
are integrated out. The width of the time window is chosen to
be order of the longest single reaction time scale, $10^2~\textrm{s}$,
which we find suitable for our parameter choice, 
by comparing the result with manual assignments.
We identify the maximum and the minimum points within each (non-overlapping)
time window and use their time series to locate the oscillation peaks 
and valleys.
Oscillation periods are then the time intervals between two successive 
peak positions, and amplitudes are the difference
in protein numbers at the valley and the following peak position
(Fig.~4(a)).
Distributions of oscillation periods and amplitudes determined
in this way are calculated from time series with a typical length
$10^{7}~\textrm{s}$.

\begin{figure}[b]
\centerline{\epsfxsize=7cm \epsfbox{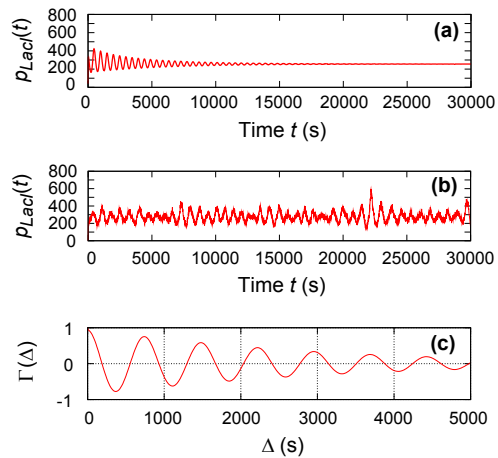}}
\caption{(Color online) Noise-induced sustained oscillations.
{\bf (a)} Deterministic protein number dynamics with the parameter set 
$(\tilde{\alpha}=2^1,\beta=2^6)$ shows damped oscillations which decay to
zero amplitude in the long-time limit according to the linear stability 
analysis.
{\bf (b)} Stochastic protein number dynamics with the same parameter set as in (a)
shows the existence of noise-induced sustained oscillations in that regime.
{\bf (c)} The auto-correlation function of the time series shown in (b), leading to
the stability measure $0.75$.
}
\end{figure}

\begin{figure*}
\centerline{\epsfxsize=0.8\linewidth \epsfbox{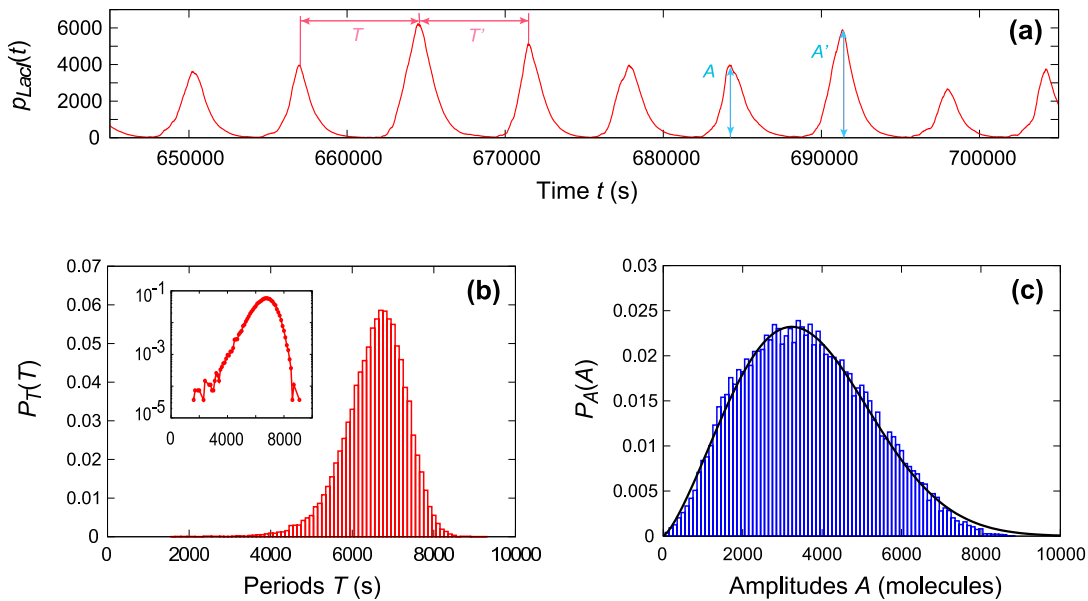}}
\caption{
(Color online) 
{\bf (a)} Portion of the sample trajectory of the LacI protein number with time
for $(\alpha,\beta)=(1/2,1/4)$.
{\bf (b)} Distributions of the oscillation periods $P_T(T)$ 
and {\bf (c)} that of the oscillation amplitudes $P_A(A)$
in the protein number oscillations obtained from the full simulation
with the same parameter set in (a).
The horizontal bin sizes are (b) 100 $s$ and (c) 100 molecules.
The inset of (b) shows $P_T(T)$ in semi-logarithmic scale,
indicating an initial exponential increase of $P_T(T)$.
The solid curve in (c) is a fit to the Weibull distribution, Eq.~(3).
}
\end{figure*}
\begin{figure*}
\centerline{\epsfxsize=\linewidth \epsfbox{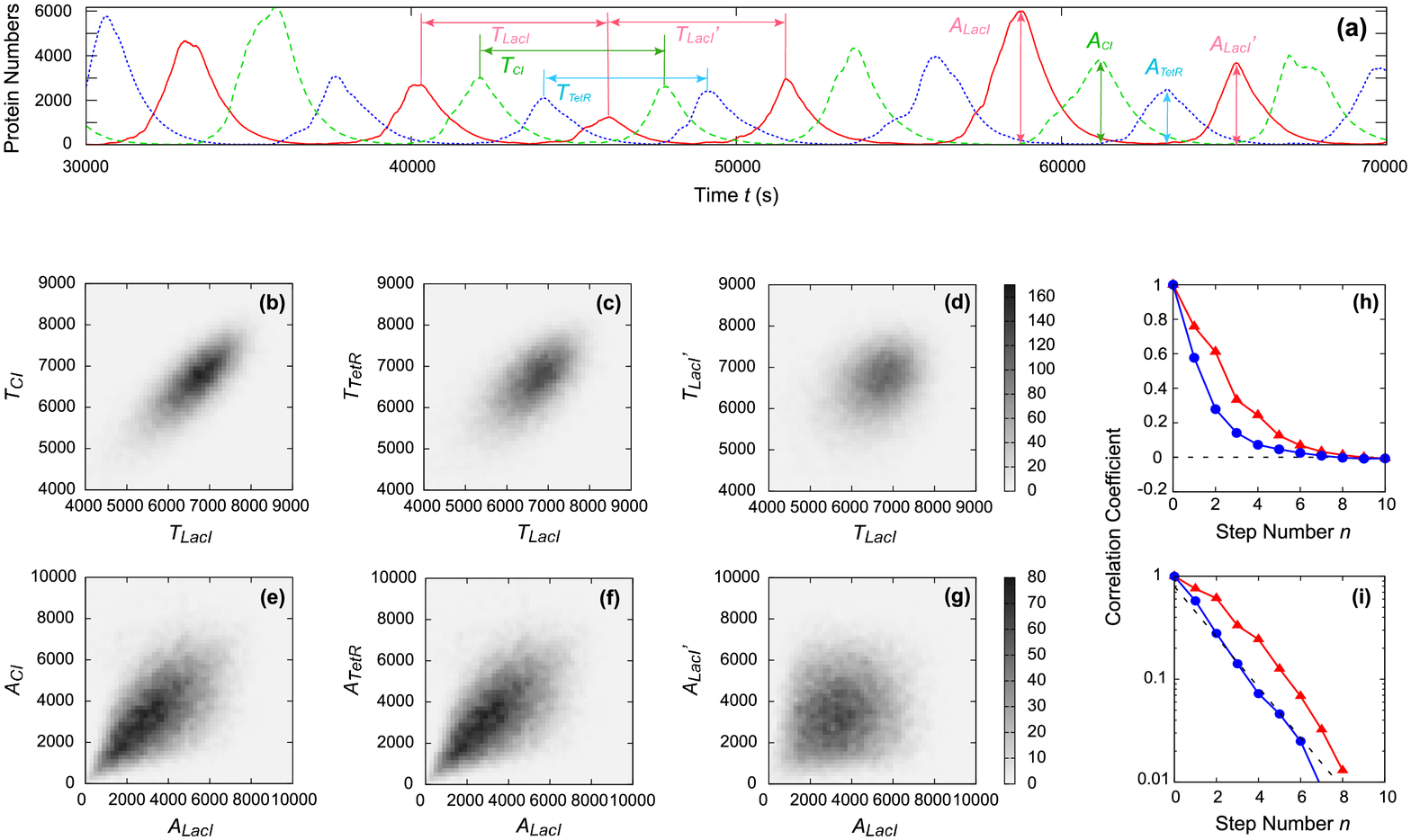}}
\caption{
(Color online) 
Noise propagation. 
{\bf (a)} Sample time series of the protein number dynamics
illustrating the correlation measurements.
{\bf (b-g)}
Correlation between the oscillation periods (b-d) and amplitudes (e-g)
along the regulatory cascade.
Plot of the consecutive $T_{CI}$ vs.\ $T_{LacI}$ (b) and 
$T_{TetR}$ vs.\ $T_{LacI}$ (c); that of two consecutive $T_{LacI}$'s (d).
Same plots are drawn for the amplitude correlation  in (e-g).
The Pearson correlation coefficients measured from the plots are
$r=0.76$ (b), $0.60$ (c), $0.31$ (d), $0.59$ (e), $0.29$ (f), and $0.15$ (g),
respectively.
{\bf (h, i)} Plots of the correlation coefficients
between oscillation periods (triangles)
and between amplitudes (circles) as functions of the regulatory
step number on a (h) linear scale and on a (i) semi-logarithmic scale.
}
\end{figure*}

\section{RESULTS}
\subsection{Stability Diagram of the Stochastic Repressilator}
The so-called stability diagram has been used to display whether the system
exhibits sustained oscillations or not in the system parameter space
\cite{repressilator,goh}.
For the stochastic dynamics, one cannot set a strict borderline
between the oscillating and the non-oscillating states due to the noisy
dynamics.  However, it is meaningful to ask if the overall
stability characteristics of the oscillatory dynamics is affected
by the stochasticity. In general, stochasticity elevates the degree
of nonlinearity in the dynamics, thereby providing an additional potential
source of instability leading to oscillatory behaviors \cite{noise-induced}.
In other cases, noise can quench the system's dynamics, thereby
destroying the oscillatory activity \cite{neuronal}.

Figure 2(a) shows the stability diagram corresponding to
stochastic oscillation of the repressilator,
plotting the stability measure, the first peak value of $\Gamma(\Delta)$, 
in $(\tilde{\alpha},\beta)$ space (Figs.~2(b,c)). 
There is a clear dark region near the center of the
stability diagram 
with decreasing stability away from the center.
This central region corresponds to the oscillation parameter region 
obtained in the deterministic modeling \cite{repressilator,goh}.
There is, however, an additional dark region near
$(\tilde{\alpha}\approx2^1, \beta\approx 2^6)$ with
as high stability measure as the central one,
suggesting the existence of sustained oscillations 
due to the fluctuation-driven instability (Fig.~3), an example
of the so-called deviant effects \cite{deviant}.
{In this additional region, which is on the border of
the sustained oscillation regime in the deterministic dynamics, 
large amplitudal fluctuations are expected.}
This additional oscillatory region, however, although theoretically 
intriguing, may not be physiologically
relevant because it corresponds to extremely unstable
proteins with very short half-lives. In typical cells,
the protein half-life is known to be 
typically much longer than that of mRNA, thus excluding
large values of $\beta$.

In the following, we fix the parameters to be
$\alpha=1/2$ and $\beta=1/4$ (indicated by $\times$ in Fig.~2(a)), 
well inside the sustained oscillation region with
the stability measure of $0.63$,
to study the fluctuation properties of the stochastic oscillations
in the system.
These parameter values are chosen to be close to
those used specifically in Ref.~\cite{repressilator},
as they reproduce well the {\it in-vivo} repressilator dynamics.
{We also verified that in this regime the stochastic simulation 
reproduces well the on-average characteristics of the deterministic dynamics
such as the mean period and the mean amplitude, allowing
us to focus exclusively on the additional role of stochasticity.}

\subsection{Single Protein-Species Level Fluctuation: Distribution of Oscillation Periods and Amplitudes}

Due to the topological symmetry of the repressilator system
and our choice of identical reaction parameters for all three components,
the statistical properties of the protein number dynamics of
the three genes are identical. Thus, we focus on the dynamics of
the first gene product, LacI. A portion of the sample trajectory
of the LacI time series is shown in Fig.~4(a).
The distribution of oscillation periods $P_T(T)$ and
that of amplitudes $P_A(A)$ obtained from the full simulation
are shown in Figs.~4(b,c).
The LacI protein number oscillates in time with a mean period
$\mu_T=6612$ $\textrm{s}$ and a standard deviation $\sigma_T=747$. 
Thus, the coefficient of variation (CV) is 
\be CV(T)\equiv\frac{\sigma_T}{\mu_T}=0.11. \ee
Looking closer, $P_T(T)$ decays like a Gaussian for large $T$
while the distribution is skewed leftward and increases as 
$\propto e^{T/T_0}$ for small $T$ (inset of Fig.~4(b)).

For the amplitude, the fluctuation is found to be stronger,
as can readily be seen from the sample time series in Fig.~4(a). 
The mean amplitude is $\mu_A=3659$ molecules, 
and the standard deviation is $\sigma_A=1588$, corresponding to a
CV of $CV(A)=0.43$, almost fourfold larger than the periods.
This higher degree of amplitude fluctuation suggests
the necessity for regulatory mechanisms controlling 
the oscillation amplitude to build a more robust cellular oscillator.
The distribution of oscillation amplitudes $P_A(A)$ is found to
be reasonably described by the Weibull distribution
\be 
P_A(A)=\frac{k}{\lambda}\left(\frac{A}{\lambda}\right)^{k-1}e^{-(A/\lambda)^k},
\ee
with parameters $k\approx2.33$ and $\lambda\approx4118$ (Fig.~4(c)).
The Weibull distribution appears in a wide range of problems,
from the extreme statistics \cite{gumbel} to the return-interval 
distributions in natural and social phenomena \cite{return} 
and fractures in disordered media \cite{fracture},
which may provide clues for the mechanisms underlying
the amplitude fluctuations.

\subsection{Noise Propagation}

Noise propagation refers to how the fluctuations
of upstream reaction processes affect the fluctuation
of the downstream processes, such as the effect
of an mRNA number fluctuation on the corresponding protein
number fluctuation \cite{paulsson1} or
that of a transcription factor number fluctuation
on its regulated protein number fluctuation \cite{propa}.
Most studies on noise propagation so far
have been concerned with a simple linear cascade \cite{propa}
or a simple autoregulated feedback structure \cite{serrano1}.

Here, we are interested in how the {\em global} noise,
such as the oscillation period and amplitude fluctuations,
propagates down the regulatory cascade.
To this end, we plot in Fig.~5 
successive oscillation periods and amplitudes for the first three steps.
For immediately following periods, the periods are well correlated
(Pearson correlation coefficient $r=0.76$; Fig.~5(b)).
The correlation decreases monotonically (Figs.~5(c,d)), but not purely exponentially 
(Figs.~5(h,i)), leading to a correlation between two consecutive periods
of the same protein number as high as $r=0.31$.
This means that the period fluctuation of a protein component
does propagate down the regulatory cascade, and that the effect is strong
enough not to be fully attenuated, not even after a complete feedback cycle.
Thus, the oscillation dynamics of the repressilator exhibits a correlated
pattern period-wise, even though the underlying reaction processes are
assumed to be purely Poissonian.

Meanwhile, the oscillation amplitude correlations (Figs.~5(e-g)) are found to 
decay purely exponentially with the regulatory step number $n$ 
as $r(n)\approx\exp(-n/n^*)$, with the characteristic step $n^*\approx1.8$,
a much faster decay than that of the oscillation periods (Figs.~5(h,i)).
This result implies that the oscillation amplitudes 
not only fluctuate more strongly individually but also are less tightly
controlled along the system. 
Amplitude control in a stochastic oscillation system
would be a more serious issue for robust functioning in this respect~\cite{autoregulatory}.

\subsection{Local Fluctuations}
Around the overall oscillatory change,
there are constant ups and downs in the protein number, 
resulting from the inherently stochastic reactions (inset of Fig.~6).
This part of the fluctuation, which we call the {\em local} noise,
can also affect the system's function as it introduces
many local maxima and minima within a global oscillation period
and provides a source of error in sensing the system's state.
The degree of local noise strongly depends on the
individual reaction rates \cite{sasai1,biham1} and, in general, anti-correlates
with the stability measure. Here, we are again interested
in the characteristics of local noise in the stable oscillation regime.

The general characteristics of the noise in protein number dynamics
are encoded in its spectral density $S(f)$, the squared magnitude of
its (complex) Fourier component with frequency $f$. 
In particular, the high-frequency 
behavior of $S(f)$ tells us about the local noise properties.
In Fig.~6, we show the spectral density of the LacI protein number.
There we can see a clear peak located at $f_1\approx1.5\times10^{-4}$ Hz
corresponding to the primary oscillation frequency, in good agreement
with the mean oscillation period in Fig.~4(b). 
This prominent peak is accompanied by additional peaks 
at integer multiples of $f_1$ with decreasing height values.
The frequency region $f\gtrsim10^{-3}$~Hz is dominated by
the local noise, in which $S(f)$ decreases algebraically with $f$ as
\be S(f)\sim f^{-2}, \ee 
implying that the protein number 
locally performs an uncorrelated random walk around the overall 
oscillatory trajectory. 
This implies that in the stable oscillation regime,
the global oscillatory dynamics of the repressilator
is primarily brought about by a system-level long-range coordination 
of the reaction processes, with only a weak perturbation
due to stochasticity.

\begin{figure}[t]
\centerline{\epsfxsize=0.85\linewidth \epsfbox{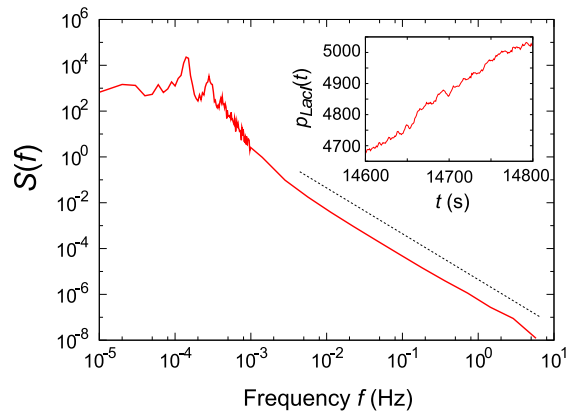}}
\caption{(Color online) Spectral density of the LacI protein number dynamics.
The peak at $f_1\approx1.5\times10^{-4}$ Hz 
is in good correspondence with the mean oscillation period.
The data for $f>10^{-3}$ are binned logarithmically for visual clarity.
The slope of the dotted straight line is $-2$, drawn for the eye.
(Inset) Zoom-up of the protein number trajectory highlighting the local 
fluctuation. }
\end{figure}

\section{Summary and Discussion}
In this paper, we have studied the noise characteristics
of stochastic (noisy) oscillations in the repressilator system
for stable oscillation regime.
Distinguishing the global and the local noises,
we first found that even in this relatively loosely regulated system,
the oscillation period is well controlled, fluctuating moderately
as approximately a Gaussian with a CV value of $0.11$, 
whereas the amplitudes fluctuate much strongly, with 
a Weibull distribution and a CV value of $0.43$.
Likewise, the period fluctuation of a protein species 
is attenuated more slowly down the regulatory cascade,
leading to a correlated behavior in adjacent oscillation
periods, a short-term collective memory, within the feedback cycle.
The amplitude fluctuation, however, propagates much more weakly,
with a characteristic step number of $1.8$, less than
the feedback cycle length. Finally, the local fluctuations
entail an inverse-square spectral density at high frequencies,
meaning that the protein number trajectory performs locally
an uncorrelated random walk around the global oscillatory dynamics.
Furthermore, we have checked that the overall noise characteristics
do not change qualitatively when we run the repressilator
for different parameter sets with varying degree of stability,
or we consider other simple oscillator circuits based on 
negative feedback, such as the 3-component cyclic circuit 
with two activators and one repressor, supporting the generality
of the results.

The repressilator is built with one of the simplest negative feedback structures
capable of sustained oscillations. Oscillatory circuits
functioning in real biological systems exhibit
much more complex structures with multiple regulatory
components, which might have evolved to ensure robust control and fine-tuning 
of the oscillatory activities. In this respect,
we hope that the current work can be extended to
systematic structural and dynamic studies towards 
the design principles of biological oscillatory networks 
including the stochastic effects \cite{novak,suel}.
Indeed, our result that the oscillation amplitude fluctuates
strongly even in the most stable oscillatory regime in the simple
circuit may imply the intrinsic tunability of the simple oscillatory
circuit \cite{kaneko} on the one hand, but at the same time, 
it demands an improved oscillatory network design for robust
and resilient oscillatory function.
Areas of future research to this end would include, {\it e.g.},
investigating the role of various regulatory mechanisms
besides simple transcription-factor binding 
for noise control, especially for the oscillation 
amplitude \cite{autoregulatory}, and the effect of a molecular network on the
stochastic dynamics \cite{goh,biham1,vilar,ghim}.
Further insights into the stochastic dynamics of the biological
oscillatory system could be gained by studies of
more complex synthetic or natural circuits \cite{hasty1,mammalian}
and of the theory of competition-based ecological systems \cite{ecol,cyclic}.

\begin{acknowledgments}
We thank the anonymous reviewer for useful comments.
This work is supported by a grant from the Korean Government (MEST) 
through KRF (KRF-2007-331-C00111). 
B.M. acknowledges the support from the Seoul Scholarship Foundation.
\end{acknowledgments}

\end{document}